\documentclass[]{spie}
\usepackage{authblk}

\usepackage{amsmath,amsfonts,amssymb}
\usepackage{graphicx}
\usepackage[colorlinks=true, allcolors=blue]{hyperref}

\usepackage[export]{adjustbox}

\title{Readout Optimization of Multi-Amplifier Sensing Charge-Coupled Devices for Single-Quantum Measurement}

\author[a,b]{Ana M.\ Botti}
\affil[a]{Fermi National Accelerator Laboratory, P.O.\ Box 500, Batavia, IL 60510, USA}
\affil[b]{Kavli Institute for Cosmological Physics, University of Chicago, Chicago, IL 60637, USA}

\author[a,c]{Brenda A. Cervantes-Vergara}
\affil[c]{Universidad Nacional Autónoma de México, Ciudad de México, México}

\author[a,d]{Claudio R.\ Chavez}
\affil[d]{Universidad Nacional del Sur (UNS), Bahía Blanca, Argentina}

\author[d,e]{Fernando Chierchie}
\affil[e]{Instituto de Investigaciones en Ingeniería Eléctrica “Alfredo C. Desages” CONICET, Bahía Blanca, Argentina}

\author[a,b,f]{Alex Drlica-Wagner}
\affil[f]{Department of Astronomy and Astrophysics, University of Chicago, Chicago, IL 60637, USA}

\author[a]{Juan Estrada}

\author[a]{Guillermo Fernandez Moroni}

\author[g]{Stephen Holland}
\affil[g]{Lawrence Berkeley National Laboratory, One Cyclotron Rd, Berkeley, CA 94720, USA}

\author[a,d,h]{Blas J.\ Irigoyen Gimenez}

\affil[h]{Facultad de Ingeniería, Universidad Nacional de Asunción, San Lorenzo, Paraguay}

\author[a,d,e]{Agustin J.\ Lapi}

\author[b,i]{Edgar Marrufo Villalpando}
\affil[i]{Department of Physics, University of Chicago, Chicago, IL 60637, USA}

\author[j,k]{Miguel Sofo Haro}
\affil[j]{Universidad Nacional de Córdoba, Instituto de Física Enrique Gaviola (CONICET)}
\affil[k]{Reactor Nuclear RA0 (CNEA), Córdoba, Argentina}

\author[a]{Javier Tiffenberg}

\author[a]{Sho Uemura}

\author[g,l]{Kenneth Lin}

\affil[l]{Department of Astronomy, University of California, Berkeley, CA 94720 USA}

\author[g]{Armin Karcher}

\author[m]{Julien Guy}
\affil[m]{University of California, Berkeley, USA}

\author[m,g]{Peter E. Nugent}

\authorinfo{Further author information: Blas J. Irigoyen Gimenez\\E-mail: blasiri16@gmail.com, Telephone: +595 982 966724}

\begin{document} 
\maketitle

\begin{abstract}
The non-destructive readout capability of the Skipper Charge Coupled Device (CCD) has been demonstrated to reduce the noise limitation of conventional silicon devices to levels that allow single-photon or single-electron counting. The noise reduction is achieved by taking multiple  measurements of the charge in each pixel. These multiple measurements come at the cost of extra readout time, which has been a limitation for the broader adoption of this technology in particle physics, quantum imaging, and astronomy applications. This work presents recent results of a novel sensor architecture that uses multiple non-destructive floating-gate amplifiers in series to achieve sub-electron readout noise in a thick, fully-depleted silicon detector to overcome the readout time overhead of the Skipper-CCD. This sensor is called the Multiple-Amplifier Sensing Charge-Coupled Device (MAS-CCD) can perform multiple independent charge measurements with each amplifier, and the measurements from multiple amplifiers can be combined to further reduce the readout noise. We will show results obtained for sensors with 8 and 16 amplifiers per readout stage in new readout operations modes to optimize its readout speed. The noise reduction capability of the new techniques will be demonstrated in terms of its ability to reduce the noise by combining the information from the different amplifiers, and to resolve signals in the order of a single photon per pixel. The first readout operation explored here avoids the extra readout time needed in the MAS-CCD to read a line of the sensor associated with the extra extent of the serial register. The second technique explore the capability of the MAS-CCD device to perform a region of interest readout increasing the number of multiple samples per amplifier in a targeted region of the active area of the device.

\end{abstract}


\section{Introduction}

Silicon semiconductor sensors with sub-photon or sub-electron noise are a major scientific breakthrough \cite{Simoen_1999, janesick_1990,boukhayma2017ultra} since they allow detecting faint signals in the silicon that were hidden by the statistical fluctuations of the electronic noise in the output amplifiers. While some of these technologies utilize signal amplification by charge multiplication \cite{Hynecek2003, BUZHAN2003}, others have demonstrated single-photon detection using small sensing structures with high charge-to-voltage gain \cite{Fossum_2016}.
A third class of detectors performs nondestructive readout to reduce the readout noise by averaging several measurements of the same collected charge. The non-destructive readout was originally enabled by a floating-gate amplifier (FGA) \cite{Wen_1974} and later proposed as the readout stage of a novel Charge Coupled Device (CCD), named the Skipper CCD \cite{Janesick_patent}.
Recent results from $p$-channel, fully-depleted Skipper CCD sensors fabricated with high-resistivity silicon \cite{Holland:2003, HV_2006, HV_2009} demonstrate that the readout noise can be reduced to arbitrarily low levels \cite{skipper_2012, Tiffenberg:2017aac, cancelo2021low} with extremely low dark current \cite{cababie_2022, Barak2020} and high quantum efficiency for blue to near-infrared photons \cite{Drlica_2020}. 

The main drawback of the Skipper CCDs is their slow readout speed at the desired noise level.  In recent years, many applications such as dark matter searches \cite{Barak2020, OSCURA2020}, neutrino detection \cite{violeta2020}, and studies of fundamental properties of silicon \cite{Rodrigues_2020, botti_2022} have exploited the capability of Skipper CCDs despite this caveat; however, other applications such as quantum imaging \cite{estrada2021ghost}, astronomical instrumentation \cite{Drlica_2020, RauscherNASA2019}, and sub-shot-noise microscopy \cite{Samantaray2017}, cannot fully profit from the Skipper CCD due to the long readout time. Recent works \cite{holland_2023, MASCCD8_2024} presented a new readout architecture that leverages multiple non-destructive readout stages to reduce readout noise which allows achieving single-quantum measurements in a shorter time compared with the Skipper-CCD. The technology is called Multi-Amplifier Sensing Charge Coupled Device (MAS-CCD) \cite{holland_2023} and has been identified as a promising candidate for optical light applications \cite{MASCCD8_2024,MASCCD16_2024,lin2024}.  In this manuscript, we explore slightly different operation schemes that enhance the readout operation in terms of efficient use of the readout time and noise reduction. 
The second section brings the main features of the architecture that will be used in the later sections. The third section focused on a continuous readout operation where the information from different rows of the sensor are simultaneously in the serial register to time efficiently read the device. The fourth section explored the idea of region of interest to select regions of the device which are read out using different number of measurements per pixel in each amplifier. The later technique was explored in the past in Skipper-CCDs \cite{chierchie_2021} and its use is extended in this case for MAS-CCDs.

\section{Detector Architecture}
\label{sec:architecture}

\begin{figure*}[ht!]
    \centering
    \includegraphics[width=1\textwidth]{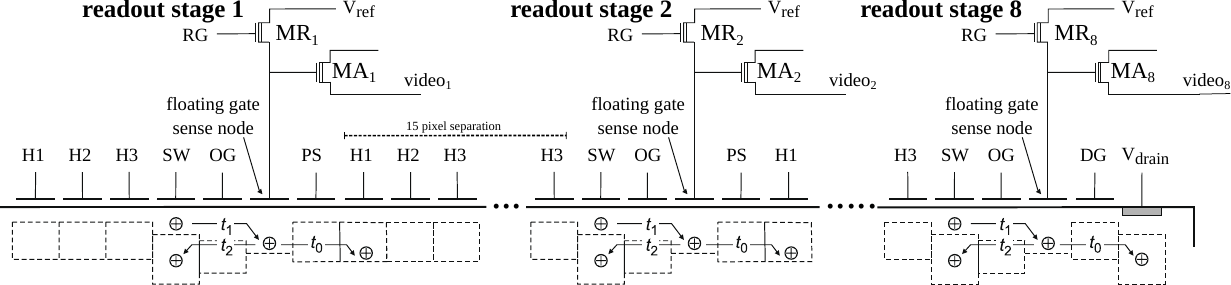}
    \caption{Architecture of the eight inline amplifiers at the end of the serial register of the MAS-CCD. The dashed line shows a qualitative potential value in the channel generated by the clocks applied to the gates to move the charge through the different structures. Figure taken from \cite{MASCCD8_2024}. }
    \label{fig:architecture}
\end{figure*}

The MAS-CCD \cite{MASCCD8_2024} uses a series of output amplifiers in its readout  capacitively connected to the sensor channel (Fig.~\ref{fig:architecture}) via a floating gate (FG), which allows for non-destructive charge measurements as in the Skipper CCDs. 
Furthermore, the MAS setup also enables charge transport through this output register, typically called the serial register, without degradation, so the pixel charge can be measured multiple times by multiple amplifiers (MA$_1$, $\dots$, MA$_8$ in Fig.~\ref{fig:architecture}) in each output stage. 

During normal operation, each amplifier simultaneously measures the charge packets from different pixels.

The final pixel value is calculated by averaging the available samples from each of the amplifiers.

\begin{equation}
    \text{pixel value} = \frac{1}{N_{a}}\frac{1}{N_s}\sum^{N_{a}}_{j=1}\sum^{N_{s}}_{i=1}s_{j,i},
    \label{eq: combination}
\end{equation}

\noindent where $s_{j,i}$ is the charge measurement sample $i$ from the amplifier $j$, $N_{a}$ is the number of amplifiers in the serial register, and $N_s$ is the number of samples taken with each amplifier. Assuming that the readout noise of each amplifier is independent and similar in standard deviation ($\sigma_0$), the standard deviation of the readout noise in the final measurement is:
\begin{equation}
\label{eq:noise reduction}
    \sigma = \frac{\sigma_0}{\sqrt{N_s}\sqrt{N_{a}}},
\end{equation}
which has an additional reduction factor ($\sqrt{N_a}$) compared to the Skipper-CCD. 

For the results presented in this article two MAS-CCD with eight and sixteen channels were employed. 
Under the auspices of the U.S. Department of Energy (DOE) Quantum Information Science initiative, $p$-channel 675$\mu$m-thick MAS-CCDs were fabricated on high-resistivity $n$-type silicon (${\sim}\,10$ k$\Omega$\,cm) with eight and sixteen amplifiers. 
The sensors were designed at LBNL to be operated as thick fully-depleted devices with high quantum efficiency over a broad wavelength range \cite{Holland:2003} (see \cite{holland_2023} for more details on the fabricated sensors). The sensors were fabricated at Teledyne DALSA Semiconductor, diced at LBNL, and packaged/tested at Fermilab. The results presented in section \ref{sec:continuous readout} comes from a front-illuminated fully depleted device with sixteen channels in the output stage. The results presented in section \ref{sec:roi} comes from an eight channel sensor fabricated following a hybrid fabrication production model used for DECam and DESI CCDs \cite{HOLLAND2007653}, where specialized processing of the back surface and metalization was done at LBNL MicroSystems Laboratory to produce 250$\mu$m thick back-illuminated wafers. For both sensors the pixel pitch is 15\,$\mu$m the output amplifiers are separated by $k_{\text{iAmp}} = 15$ pixels, and both have $k_{\text{ex}} = 27$ extended pixels (pre-scan pixels), used to transfer the charge packets from the serial register to the output stages.

\section{Continuous readout}
\label{sec:continuous readout}

\begin{figure}[h!]
    \centering
    \includegraphics[width=0.8\textwidth]{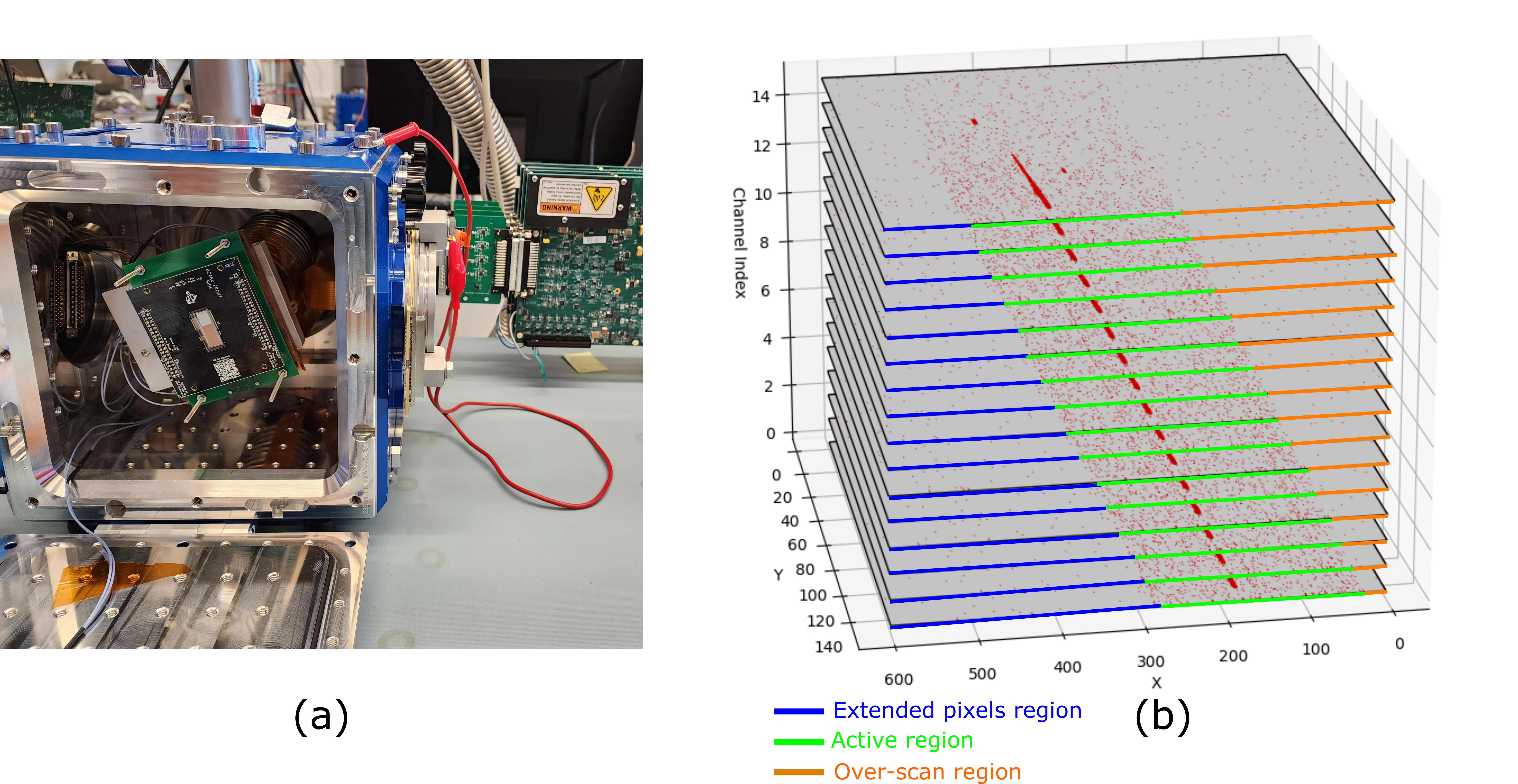}
    \caption{a) Experimental setup of the 16-channels MAS-CCD. b) Stack of the output images obtained from each output amplifier of a 16-channel MAS-CCD using the typical readout scheme. The color indicates each region of the output images taken by each amplifier.}
    \label{fig:previous image scheme}
\end{figure}

In the operation modes used to read this type of sensor in \cite{MASCCD8_2024,MASCCD16_2024} the entire row in the active region of the sensor is dumped in the serial register. Then the charge is moved across it through the output amplifiers as shown in Fig. \ref{fig:architecture}. The next row in the active region is moved to the serial register after the last pixel of the previous line reaches the last amplifier of the chain. With this process, the output images for the different channels look like the ones presented in Fig. \ref{fig:previous image scheme}. The active region is the portion located in the center of each image, marked with a green line. The red tracks are ionization produced in the active volume. The pixels on the left of this region (marked with a blue line) correspond to the pre-scan of each amplifier. Its width represents the distance in pixels of the serial register between each stage and the active region ($k_{\text{iAmp}}(j-1) + k_{\text{ex}}$ for the output stage $j$). Therefore, it increases for amplifiers located farther from the active region. The region on the right (marked with an orange line) is the virtual arbitrary overscan for each amplifier. The same ionization tracks are observed in all amplifiers. In this readout operation, a significant amount of time is wasted reading pixels that are not part of the active region. 

A faster readout scheme can be used if, during the readout operation, a row of the active region is dumped into the serial register just after the first $N_\text{col}$ pixels (The number of pixel columns in the active region) in the serial register are empty, even if some of the pixels of the previous row are still being read in the output stages.
In this way, there is no time consumed in measuring the pre-scan pixels. Only the pre-scan pixels of the first row of the sensor contribute to the readout overhead. 

\begin{figure*}[h!]
    \centering
    \includegraphics[width=\textwidth]{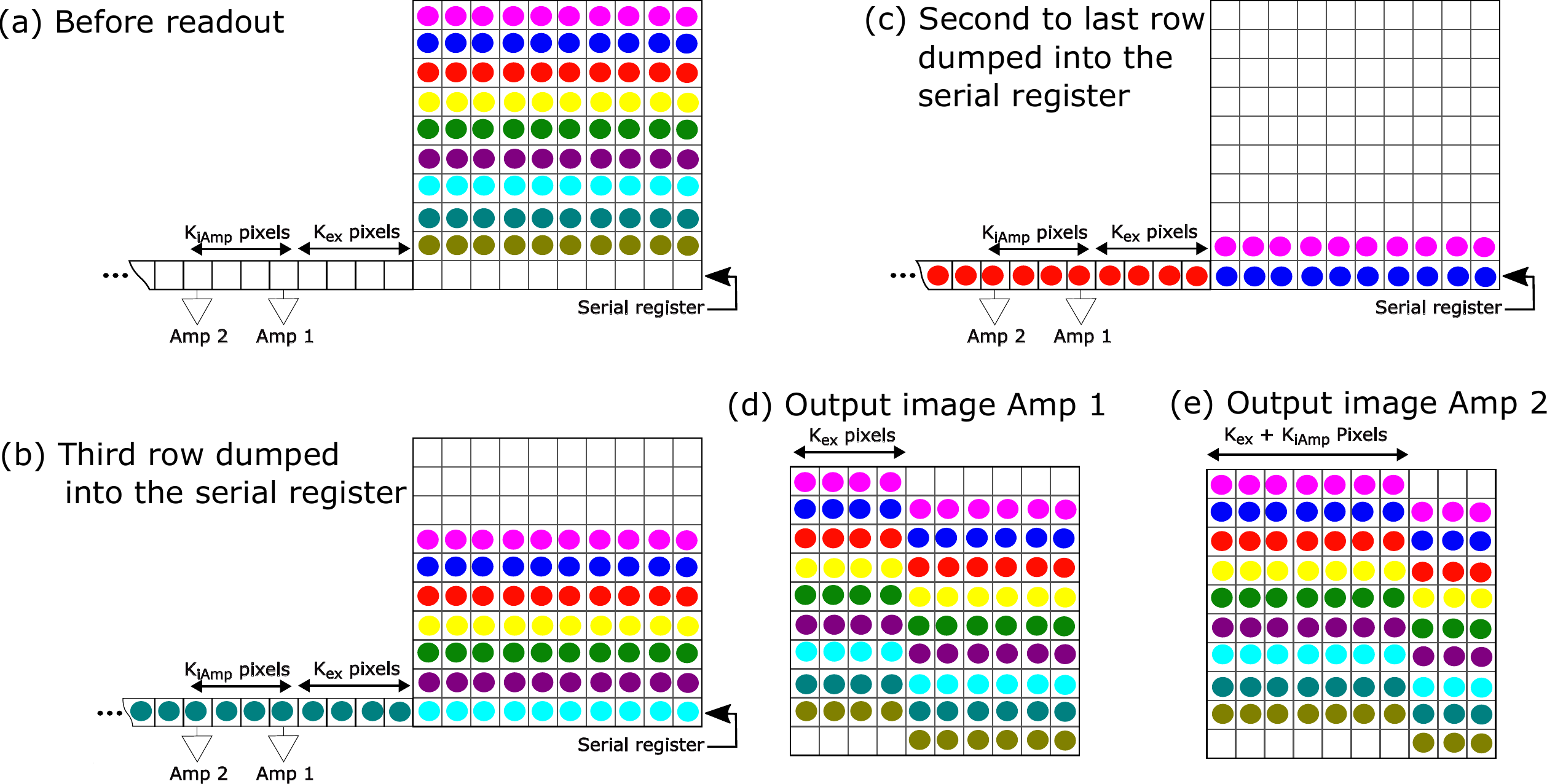}
    \caption{The process steps of the continuous readout scheme are shown in figures (a), (b), and (c). (e) and (f) representation of our output image using the method.}
    \label{fig:cont_read_sch}
\end{figure*}

This readout strategy is summarized in the successive CCD cartoons in Fig.\ref{fig:cont_read_sch}. Figure \ref{fig:cont_read_sch}(a) shows the pixels in the active region with a different color for the collected charge on each row. At this time the serial register is empty. The readout starts by dumping the first row into the serial register and moving the charge towards the output amplifiers. Once the pixels in the serial register corresponding to the active regions are empty, the second row is dumped is the serial register. The process is repeated for the following rows. 
 
In particular, Fig.\ref{fig:cont_read_sch}(b) shows the status of the serial register when the third row is dumped. In the serial register, there are no empty pixels between the pixels from each row. Figure \ref{fig:cont_read_sch}(c) displays the moment when the penultimate row of the active region is dumped in the serial register. 

The output images as they are arranged by our readout system (the Low Threshold Acquisition \cite{cancelo2021low}) are shown for completeness in Fig.\ref{fig:cont_read_sch}(d) and (e) for the first and second amplifiers, respectively.
The first row in each image, contains empty pixels which corresponds to the pre-scan pixels for each amplifier, that represents the amount of pixels between the serial register and the corresponding  j-th amplifier ($k_{\text{iAmp}}(j-1) + k_{\text{ex}}$) pixels. The images should be rearranged to map the pixel information to the right position in the array.

Then, all the pixels read by all the amplifiers are active pixels. This also allows for a virtual over-scan with the same number o pixels for the image from each channel. 

\begin{figure*}[h!]
    \centering
    \includegraphics[width=0.8\textwidth]{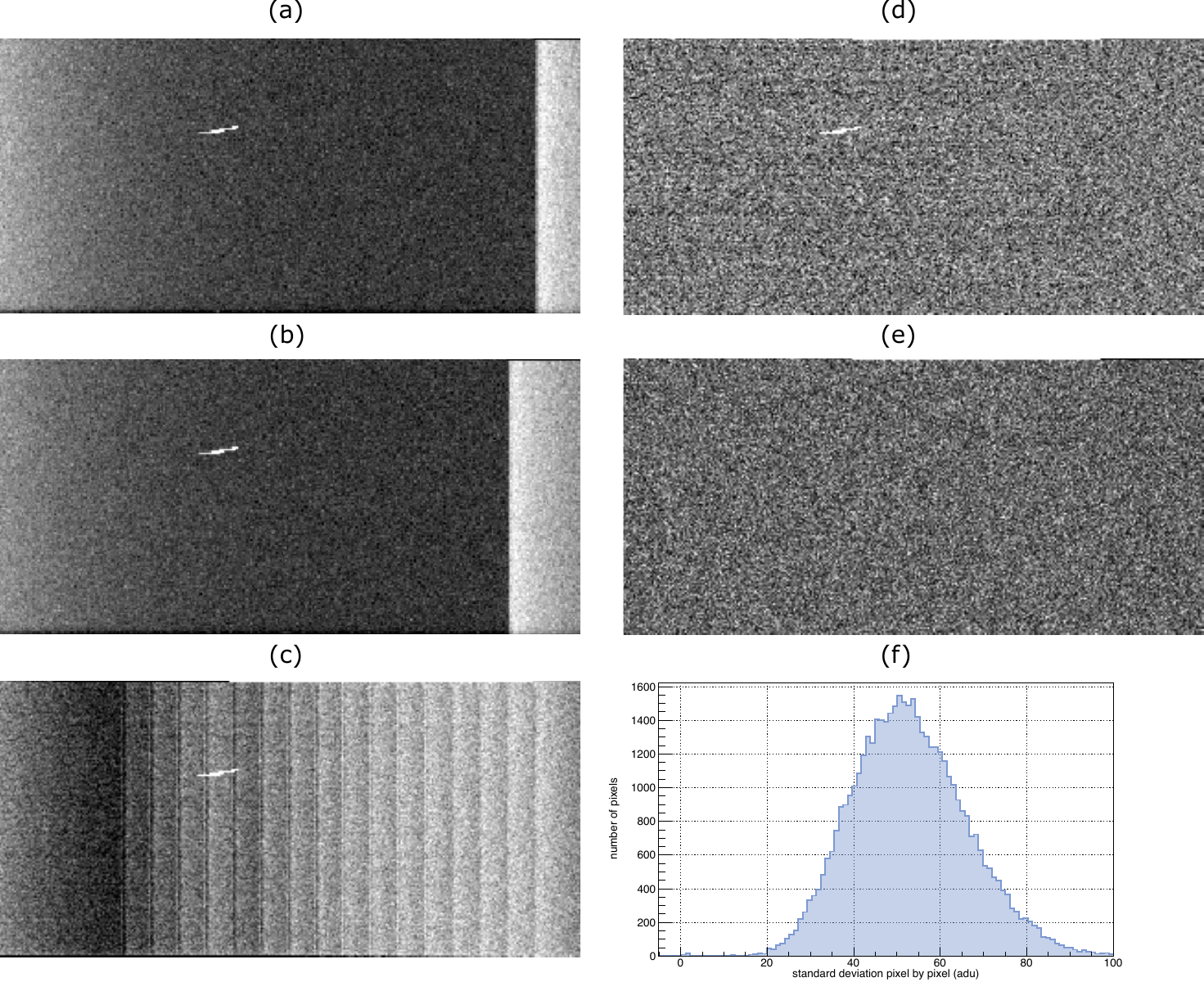}
    \caption{a) Image obtained from the first amplifier with the continuous readout method. b) Image obtained from the second amplifier with the continuous readout method. c) Image from the first amplifier after the rearrangement of the pixels. d) Image from the second amplifier after the rearrangement of the pixels. d) Image corrected by subtracting a median image of many dark rearranged images. e) Median image calculated from 20 dark images taken, after the rearrangement of the pixels and combining each channel information for each image. f) Histogram of the standard deviation of many corrected images pixel by pixel.}
    \label{fig:cont_readout}
\end{figure*}

An image obtained with the continuous readout, after pixels rearrangement is shown in Fig. \ref{fig:cont_readout}(a). 

A more illuminated strip can be seen on the right side of the image with a different baseline due to the effect of the vertical clocking before each row is completely read. This strip has a width of $k_{ex}$ columns of pixels and increases towards the left by ${k_{i}}_{Amp}$ columns of pixels for each channel, as shown for the second amplifier in Fig. \ref{fig:cont_readout}(b).

By mixing the data from the sixteen channels, the image shown in Fig. \ref{fig:cont_readout}(c) is obtained, where the baseline changes due to the vertical clocks for all the channels are clearly visible.

One of the key points for the feasibility of this technique is that this new baseline contribution can be subtracted from the images, in this case, using a bias image. For this purpose, 20 dark images are taken with single sample per pixel, and the information from the different channels are combined using equation \ref{eq: combination}. A general single offset in the image (calculated in the overscan region of the image) is subtracted from each of the combined images. Subsequently, the a stacked of combined images with no offset are used to calculate the bias image pixel by pixel. The bias image is then used to subtract the baseline structure from the data of interest. An example of this resultant image after this processing is shown in Fig. \ref{fig:cont_readout}(c) where no evident transition in the baseline properties is observed.

To further evaluate the statistical properties,  a stack of many corrected images is used to calculate the standard deviation of each pixel of the output images. This values are represented in first as an image shown in Fig. \ref{fig:cont_readout}(e).  In this case the image shows in gray scale the standard deviation of each pixel showing no evident increase on the standard deviation specially in the places where the vertical transitions are observed in Fig.                       \ref{fig:cont_readout}(c).  In the gray color scale white indicates larger standard deviation. At the same time, the  calculated individual standard deviation of each pixel can be grouped in a histogram, shown in \ref{fig:cont_readout}(f) with a mean around 50 ADU, which correspond to one electron of noise as expected for this sensor after combining the information from the sixteen channels. The pixels were readout at a rate of 70,000 pixels per second using an integration time of 3.6 $\mu$s.

\section{Sub-electron noise within a Region of Interest (ROI)}
\label{sec:roi}

\begin{figure}[h!]
    \centering
    \includegraphics[width=0.8\textwidth]{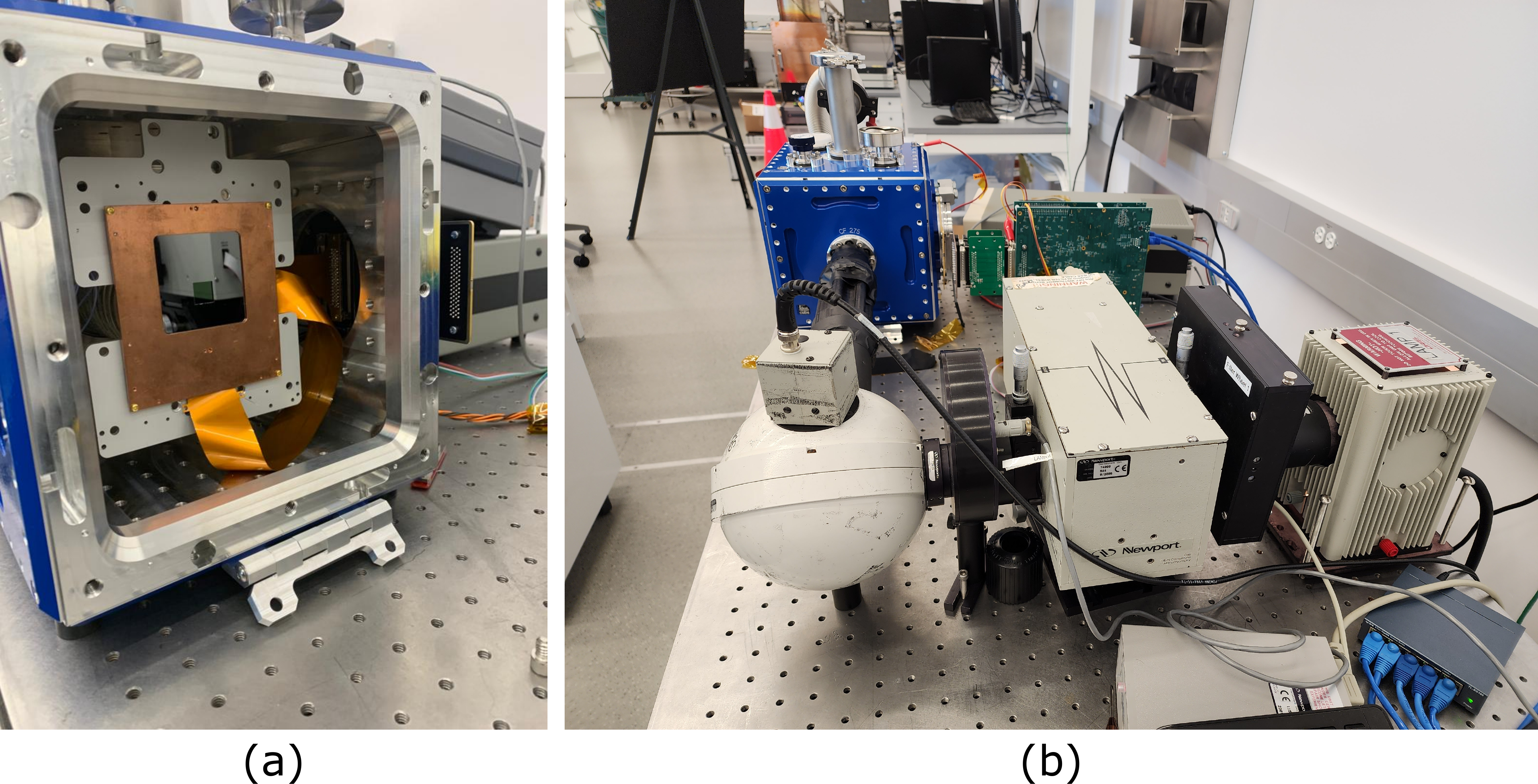}

    \caption{Setup used to test the region of interest readout using an 8-channels MAS-CCD. a) Picture of the inside of the dewar showing the sensor. b) Picture of the setup with the optical lighting system, consisting of a lamp, a filter wheel, a monochromator, a concentrator sphere and a calibrated photodiode to measure the light power.}
    \label{fig:setup}
\end{figure}

Another readout strategy explored in this work is related to the definition of regions of interest (ROI) where the number of non-destructive readouts is increased in pixels where a lower noise is required to detect the low light levels. The idea was explored in the past for the Skipper-CCD \cite{chierchie_2021} and here is extended to the MAS-CCD where the same information from different amplifiers can be combined. A different CCD testing setup (shown in Fig. \ref{fig:setup}a) that allows light to be projected onto the sensor was used for these tests.

\begin{figure*}[h!]
    \centering
    \includegraphics[width=0.7\textwidth]{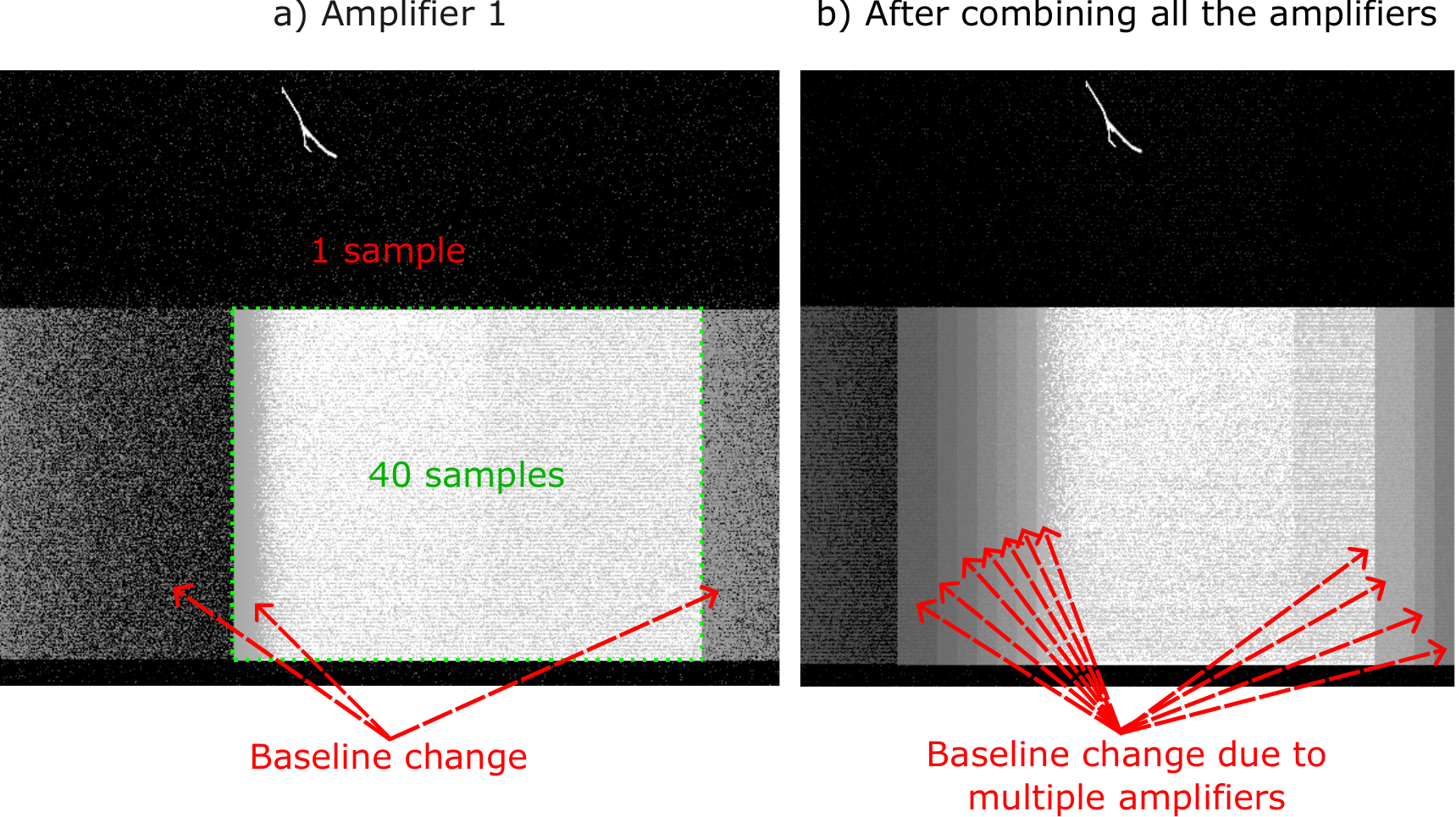}
    \caption{Image taken using ROI. (a) Image from a single amplifier. (b) After combining the information from seven amplifiers.}
    \label{fig:ch1_mix_smart_readout}
\end{figure*}

For this test, a region of 360 by 270 pixels were read using 40 non-destructive measurements per amplifier while the rest of the sensor is read out using single sample per amplifier. The achieved noise in the ROI is compared to the expected reduction when the entire image is read out at the same number of samples. Figure \ref{fig:ch1_mix_smart_readout}a shows the output image of one of the sensor channels using this readout mode. The central rectangular region highlighted by the dashed lines has the 40 non-destructive samples per pixels and the rest of the sensor is read out using a single sample per pixel. One of the drawbacks in changing the readout sequence is the movement in the pixels baseline due to the change in the clock sequence controlling the sensor. Two effects are evident from the figure. Firstly, each region has a different offset level and, secondly, this offset varies in the transitions between regions. This effect is similar to the change in the baseline produced by the vertical clocks during normal readout operation mentioned in the previous section. This mechanism, in particular for the ROI readout with non-destructive readout sensor has been explored in the past \cite{chierchie_2021}. This effect appears in all the channels. The aim of this work is to demonstrate that when all the information from all the channels is combined, the effect can be subtracted without leaving residues that could impact the noise performance in the ROI. 

Figure \ref{fig:ch1_mix_smart_readout}b shows the output image after combining the information from all the channels as they come out from the CCD controller software. Since the sensor amplifiers perform the multiple sampling per pixel at the same time, they actually evaluate different portions of the pixel matrix. This is evident in the figure thanks to the different offsets in the smart regions due to the multiple amplifiers. The individual ROI regions intersect in the middle, so that these pixels are measured with the maximum precision by all the amplifiers. The separation between the regions from each amplifier corresponds to the physical spatial separations of the amplifiers in the output stages chain. In this particular sensor with 8 output transistors, the video signal from the seventh amplifier was not giving any charge signal, so we decided to directly exclude the channel from the analysis without performing a thorough debugging of the electronic chain to identify the source of the problem.

In order to eliminate the baseline fluctuations in the different regions, a set of dark images was acquired to compute a bias image using the median operation. The steps used in the process are: all the available measurements (from the same amplifier and from different amplifiers) of the pixels are averaged using equation \ref{eq: combination} into a single image. This produces a single output image similar to Fig. \ref{fig:ch1_mix_smart_readout}b for each exposure. Then, a single offset (calculated as the mean in the overscan region of the sensor) from the combined image is subtracted. The resulting images from this step are stacked to calculate a bias image using the median value of each pixel.

\begin{figure*}[h!]
    \centering
    \includegraphics[width=0.7\textwidth]{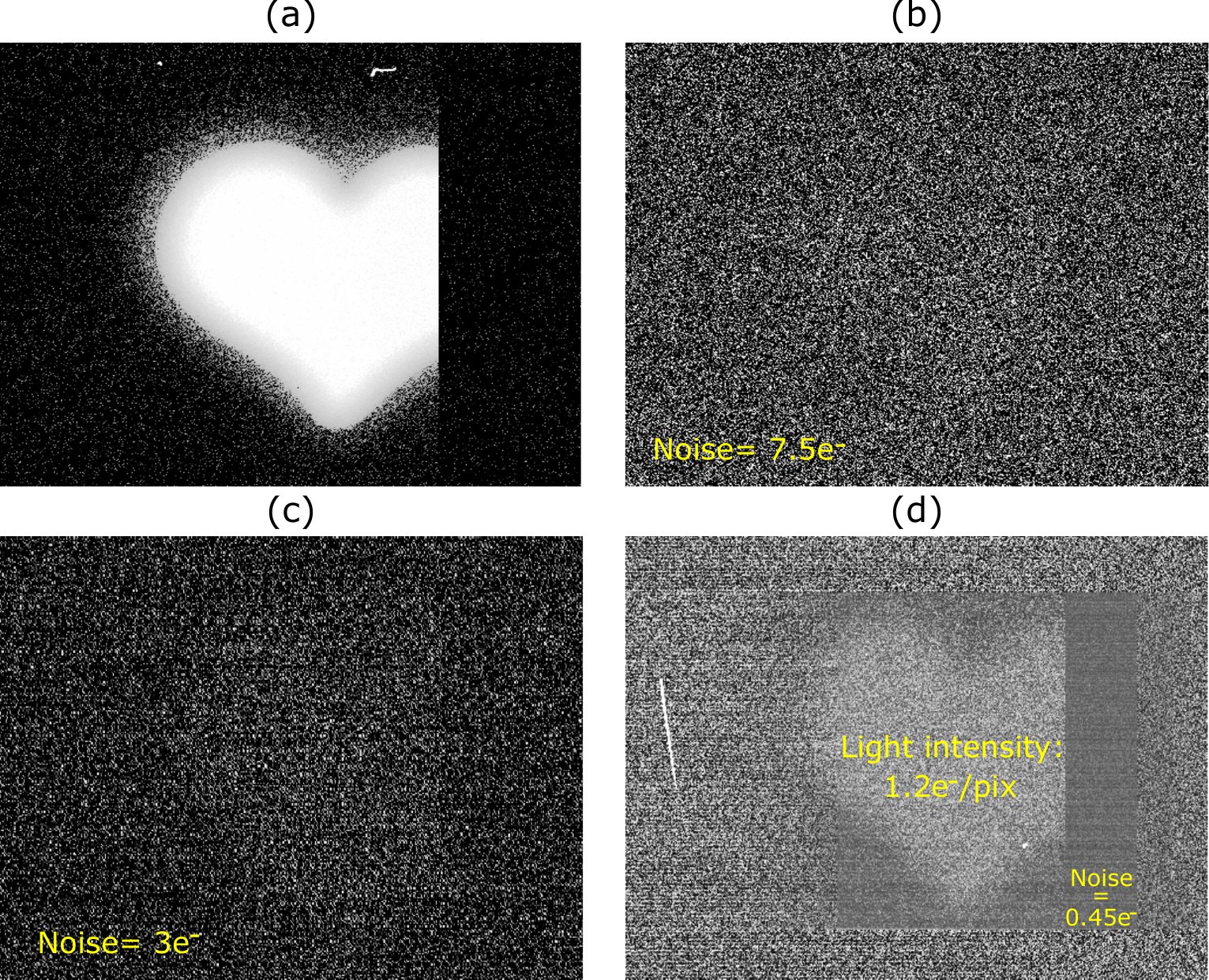}
    \caption{Same region of a 8-channels MAS-CCD taken in different modes. (a) Image from a single amplifier taken after projecting a high-intensity light signal from a known object onto the CCD. This image was used to define the ROI geometry. (b), (c), (d) Images after projecting the same region with a low light level of around 1.2e- per pixel from photons with a wavelength of 500 nm and using a single amplifier with a single sample per pixel (b), the combination of all the available amplifiers with a single sample per amplifier (c), and with the definition of ROI with 40 samples per pixel (d). }
    \label{fig:smart_readout}
\end{figure*}

The bias image calculated following the previous procedure is then subtracted from the images of interest. To make the result more attractive to the reader, we projected a known object to the sensor with a light intensity smaller than the noise fluctuations using a single sample per pixel. The light collection was first tested using a large exposure to find its location and shape on the sensor (Fig.\ref{fig:smart_readout}a) which shows the "heart" shape observed in this case.  The missing part of the object is due to misalignment of the projection of the pattern with the observed active region in the sensor. This image was used to define the ROI explained before. The light intensity and the exposure time were then reduced to produce an average signal around 1.2e- per pixel. The results from these exposures in the same region of the sensor are shown in Fig. \ref{fig:smart_readout}b, c and d. Figure \ref{fig:smart_readout}b shows the image observed with a single amplifier using a single sample per pixel for the entire sensor. In this case, the noise does not allow to observe the projected object with low intensity. When the information of the available amplifiers are combined (still using a single sample per amplifier), Fig. \ref{fig:smart_readout}c, the impression of a whiter shadow seems to appear in the central part of the image, but still the object is not clearly observable. The last test consists of using the ROI with 40 samples per pixel. At this point, the projected image is clearly visible thanks to the low noise achieved, of around 0.45e-. Another interesting aspect is that there are no evident residuals of the baseline fluctuations in the final image. Moreover, the noise in the ROI is similar to the  expected from the equation \ref{eq:noise reduction} for $N_s = 40$, $N_a=7$ starting from a single sample noise per amplifier of 7.5e-. Although the integration time in the pixel readout is the same as in the previous section, the noise performance of this setup is inferior due to slower electronic of the readout system.

\newpage

\section{Conclusions}

A new continuous readout method for a novel multiple-amplifier architecture silicon detector, known as the MAS-CCD, was presented, This method avoids unnecessary time loss in the readout process and has proven to have no negative impact on the readout noise.

Furthermore, sub-electron readout noise within a Region of Interest was demonstrated. This performance was achieved with a smart readout, which saves readout time by focusing only on taking multiple non-destructive samples per pixel in the region of the image where the target object is located.
The effects of these methods on the readout process outcome has been thoroughly described, confirming the absence of any negative contribution to the readout noise.
These methods will help reduce the readout times presented in previous works with this type of sensors for the same level of readout noise performance.

\section* {Acknowledgments}

The fully depleted Skipper-CCD was developed at Lawrence Berkeley National Laboratory, as were the designs described in this work. The CCD development work was supported in part by the Director, Office of Science, of the U.S. Department of Energy under No. DE-AC02-05CH11231.
The fabrication of the MAS-CCD was partially funded by Javier Tiffenberg through the DOE Early Career Award (DOE-ECA).
This research has been sponsored by the Laboratory Directed Research and Development Program of Fermi National Accelerator Laboratory (L2019-011, L2022.053), managed by Fermi Research Alliance, LLC for the U.S. Department of Energy.
This research has been partially supported by Guillermo Fernandez Moroni's DOE Early Career research program.
Support was also provided by NASA APRA award No.80NSSC22K1411 and a grant from the Heising-Simons Foundation (\#2023-4611).

\bibliography{bibliography} 

\begin{thebibliography}{10}

\bibitem{Simoen_1999}
E.~Simoen and C.~Claeys, ``On the flicker noise in submicron silicon mosfets,''
  {\em Solid-State Electronics}~{\bf 43}(5), pp.~865--882, 1999.

\bibitem{janesick_1990}
J.~R. Janesick, T.~S. Elliott, A.~Dingiziam, R.~A. Bredthauer, C.~E. Chandler,
  J.~A. Westphal, and J.~E. Gunn, ``{New advancements in charge-coupled device
  technology: subelectron noise and 4096 x 4096 pixel CCDs},'' in {\em
  Charge-Coupled Devices and Solid State Optical Sensors},  M.~M. Blouke, ed.,
  {\bf 1242}, pp.~223 -- 237, International Society for Optics and Photonics,
  SPIE, 1990.

\bibitem{boukhayma2017ultra}
A.~Boukhayma, {\em Ultra Low Noise CMOS Image Sensors}, Springer Theses,
  Springer International Publishing, 2017.

\bibitem{Hynecek2003}
J.~Hynecek and T.~Nishiwaki, ``Excess noise and other important characteristics
  of low light level imaging using charge multiplying ccds,'' {\em IEEE
  Transactions on Electron Devices}~{\bf 50}(1), pp.~239--245, 2003.

\bibitem{BUZHAN2003}
P.~Buzhan, B.~Dolgoshein, L.~Filatov, A.~Ilyin, V.~Kantzerov, V.~Kaplin,
  A.~Karakash, F.~Kayumov, S.~Klemin, E.~Popova, and S.~Smirnov, ``Silicon
  photomultiplier and its possible applications,'' {\em Nuclear Instruments and
  Methods in Physics Research Section A: Accelerators, Spectrometers, Detectors
  and Associated Equipment}~{\bf 504}(1), pp.~48--52, 2003.
\newblock Proceedings of the 3rd International Conference on New Developments
  in Photodetection.

\bibitem{Fossum_2016}
E.~R. Fossum, J.~Ma, S.~Masoodian, L.~Anzagira, and R.~Zizza, ``The quanta
  image sensor: Every photon counts,'' {\em Sensors}~{\bf 16}(8), 2016.

\bibitem{Wen_1974}
D.~Wen, ``Design and operation of a floating gate amplifier,'' {\em IEEE
  Journal of Solid-State Circuits}~{\bf 9}(6), pp.~410--414, 1974.

\bibitem{Janesick_patent}
J.~R. Janesick, ``Ultra low-noise charge coupled device,'' 1993.

\bibitem{Holland:2003}
S.~E. Holland, D.~E. Groom, N.~P. Palaio, R.~J. Stover, and M.~Wei, ``Fully
  depleted, back-illuminated charge-coupled devices fabricated on
  high-resistivity silicon,'' {\em IEEE Transactions on Electron Devices}~{\bf
  50}, pp.~225--238, Jan 2003.

\bibitem{HV_2006}
S.~E. Holland, C.~J. Bebek, K.~S. Dawson, J.~H. Emes, M.~H. Fabricius, J.~A.
  Fairfield, D.~E. Groom, A.~Karcher, W.~F. Kolbe, N.~P. Palaio, N.~A. Roe, and
  G.~Wang, ``{High-voltage-compatable fully depleted CCDs},'' in {\em High
  Energy, Optical, and Infrared Detectors for Astronomy II},  D.~A. Dorn and
  A.~D. Holland, eds.,  {\bf 6276}, p.~62760B, International Society for Optics
  and Photonics, SPIE, 2006.

\bibitem{HV_2009}
S.~E. Holland, W.~F. Kolbe, and C.~J. Bebek, ``Device design for a
  12.3-megapixel, fully depleted, back-illuminated, high-voltage compatible
  charge-coupled device,'' {\em IEEE Transactions on Electron Devices}~{\bf
  56}(11), pp.~2612--2622, 2009.

\bibitem{skipper_2012}
G.~Fernandez~Moroni, J.~Estrada, G.~Cancelo, S.~Holland, E.~Paolini, and
  H.~Diehl, ``Sub-electron readout noise in a {Skipper} {CCD} fabricated on
  high resistivity silicon,'' {\em Experimental Astronomy}~{\bf 34}, 07 2012.

\bibitem{Tiffenberg:2017aac}
J.~Tiffenberg, M.~Sofo-Haro, A.~Drlica-Wagner, R.~Essig, Y.~Guardincerri,
  S.~Holland, T.~Volansky, and T.-T. Yu, ``{Single-electron and single-photon
  sensitivity with a silicon {Skipper} {CCD}},'' {\em Phys. Rev. Lett.}~{\bf
  119}(13), p.~131802, 2017.

\bibitem{cancelo2021low}
G.~I. Cancelo, C.~Chavez, F.~Chierchie, J.~Estrada, G.~Fernandez-Moroni, E.~E.
  Paolini, M.~S. Haro, A.~Soto, L.~Stefanazzi, J.~Tiffenberg, K.~Treptow,
  N.~Wilcer, and T.~J. Zmuda, ``{Low threshold acquisition controller for
  Skipper charge-coupled devices},'' {\em Journal of Astronomical Telescopes,
  Instruments, and Systems}~{\bf 7}(1), pp.~1 -- 19, 2021.

\bibitem{cababie_2022}
L.~Barak, I.~M. Bloch, A.~Botti, M.~Cababie, G.~Cancelo, L.~Chaplinsky,
  F.~Chierchie, M.~Crisler, A.~Drlica-Wagner, R.~Essig, J.~Estrada, E.~Etzion,
  G.~Fernandez~Moroni, D.~Gift, S.~E. Holland, S.~Munagavalasa, A.~Orly,
  D.~Rodrigues, A.~Singal, M.~S. Haro, L.~Stefanazzi, J.~Tiffenberg, S.~Uemura,
  T.~Volansky, and T.-T. Yu, ``Sensei: Characterization of single-electron
  events using a skipper charge-coupled device,'' {\em Phys. Rev. Appl.}~{\bf
  17}, p.~014022, Jan 2022.

\bibitem{Barak2020}
L.~Barak, I.~M. Bloch, M.~Cababie, G.~Cancelo, L.~Chaplinsky, F.~Chierchie,
  M.~Crisler, A.~Drlica-Wagner, R.~Essig, J.~Estrada, E.~Etzion, G.~F. Moroni,
  D.~Gift, S.~Munagavalasa, A.~Orly, D.~Rodrigues, A.~Singal, M.~S. Haro,
  L.~Stefanazzi, J.~Tiffenberg, S.~Uemura, T.~Volansky, and T.-T. Yu,
  ``{SENSEI}: Direct-detection results on sub-gev dark matter from a new
  {Skipper} {CCD},'' {\em Phys. Rev. Lett.}~{\bf 125}, p.~171802, Oct 2020.

\bibitem{Drlica_2020}
A.~Drlica-Wagner, E.~M. Villalpando, J.~O'Neil, J.~Estrada, S.~Holland,
  N.~Kurinsky, T.~Li, G.~F. Moroni, J.~Tiffenberg, and S.~Uemura,
  ``{Characterization of skipper CCDs for cosmological applications},'' in {\em
  X-Ray, Optical, and Infrared Detectors for Astronomy IX},  A.~D. Holland and
  J.~Beletic, eds.,  {\bf 11454}, pp.~210 -- 223, International Society for
  Optics and Photonics, SPIE, 2020.

\bibitem{OSCURA2020}
J.~Estrada, {\em Observatory of {Skipper} {CCD}s Unveiling Recoiling Atoms},
  2020 (accessed September 23, 2021).
\newblock \url{https://astro.fnal.gov/science/dark-matter/oscura/}.

\bibitem{violeta2020}
J.~C. D'Olivo, C.~Bonifazi, D.~Rodrigues, and G.~F. Moroni, ``{vIOLETA:
  Neutrino Interaction Observation with a Low Energy Threshold Array},'' in
  {\em XXIX International Conference in Neutrino Physics, poster 521},  June
  2020.

\bibitem{Rodrigues_2020}
D.~Rodrigues, K.~Andersson, M.~Cababie, A.~Donadon, A.~Botti, G.~Cancelo,
  J.~Estrada, G.~Fernandez-Moroni, R.~Piegaia, M.~Senger, M.~S. Haro,
  L.~Stefanazzi, J.~Tiffenberg, and S.~Uemura, ``Absolute measurement of the
  fano factor using a skipper-ccd,'' {\em Nuclear Instruments and Methods in
  Physics Research Section A: Accelerators, Spectrometers, Detectors and
  Associated Equipment}~{\bf 1010}, p.~165511, 2021.

\bibitem{botti_2022}
A.~M. Botti, S.~Uemura, G.~F. Moroni, L.~Barak, M.~Cababie, R.~Essig,
  E.~Etzion, D.~Rodrigues, N.~Saffold, M.~Sofo~Haro, J.~Tiffenberg, and
  T.~Volansky, ``Constraints on the electron-hole pair creation energy and fano
  factor below 150 ev from compton scattering in a skipper ccd,'' {\em Phys.
  Rev. D}~{\bf 106}, p.~072005, Oct 2022.

\bibitem{estrada2021ghost}
J.~Estrada, R.~Harnik, D.~Rodrigues, and M.~Senger, ``Ghost imaging of dark
  particles,'' 2021.

\bibitem{RauscherNASA2019}
B.~J. Rauscher, S.~E. Holland, L.~R. Miko, and A.~Waczynski, ``{Radiation
  tolerant, photon counting, visible and near-IR detectors for space
  coronagraphs and starshades},'' in {\em UV/Optical/IR Space Telescopes and
  Instruments: Innovative Technologies and Concepts IX},  A.~A. Barto, J.~B.
  Breckinridge, and H.~P. Stahl, eds.,  {\bf 11115}, pp.~382 -- 386,
  International Society for Optics and Photonics, SPIE, 2019.

\bibitem{Samantaray2017}
N.~Samantaray, I.~Ruo-Berchera, A.~Meda, and M.~Genovese, ``Realization of the
  first sub-shot-noise wide field microscope,'' {\em Light: Science {\&}
  Applications}~{\bf 6}, pp.~e17005--e17005, Jul 2017.

\bibitem{holland_2023}
S.~E. Holland, ``Fully depleted charge-coupled device design and technology
  development,'' {\em Astronomische Nachrichten}~{\bf 344}(8-9), p.~e20230072,
  2023.

\bibitem{MASCCD8_2024}
A.~M. Botti, B.~A. Cervantes-Vergara, C.~R. Chavez, F.~Chierchie,
  A.~Drlica-Wagner, J.~Estrada, G.~F. Moroni, S.~E. Holland, B.~J.
  Irigoyen~Gimenez, A.~J. Lapi, E.~M. Villalpando, M.~S. Haro, J.~Tiffenberg,
  and S.~Uemura, ``Single-quantum measurement with a multiple-amplifier sensing
  charge-coupled device,'' {\em IEEE Transactions on Electron Devices}~{\bf
  71}(6), pp.~3732--3738, 2024.

\bibitem{MASCCD16_2024}
A.~J. Lapi, B.~J. Irigoyen~Gimenez, M.~E. Gamero, C.~R. Chavez~Blanco,
  F.~Chierchie, G.~Fernandez~Moroni, S.~E. Holland, J.~Estrada, and
  J.~Tiffenberg, ``A sixteen multiple-amplifier-sensing ccd and
  characterization techniques targeting the next generation of astronomical
  instruments,'' 2024.

\bibitem{lin2024}
K.~{Lin}, A.~{Karcher}, J.~{Guy}, S.~E. {Holland}, W.~F. {Kolbe}, P.~{Nugent},
  and A.~{Drlica-Wagner}, ``{Multi-Amplifier Sensing Charge-coupled Devices for
  Next Generation Spectroscopy},'' {\em arXiv e-prints} , p.~arXiv:2406.06472,
  June 2024.

\bibitem{chierchie_2021}
F.~Chierchie, G.~F. Moroni, L.~Stefanazzi, C.~Chavez, E.~Paolini, G.~Cancelo,
  M.~S. Haro, J.~Tiffenberg, J.~Estrada, and S.~Uemura, ``Smart-readout of the
  skipper-ccd: Achieving sub-electron noise levels in regions of interest,'' in
  {\em 2021 Argentine Conference on Electronics (CAE)},  pp.~82--87, 2021.

\bibitem{HOLLAND2007653}
S.~Holland, K.~Dawson, N.~Palaio, J.~Saha, N.~Roe, and G.~Wang, ``Fabrication
  of back-illuminated, fully depleted charge-coupled devices,'' {\em Nuclear
  Instruments and Methods in Physics Research Section A: Accelerators,
  Spectrometers, Detectors and Associated Equipment}~{\bf 579}(2),
  pp.~653--657, 2007.
\newblock Proceedings of the 6th "Hiroshima" Symposium on the Development and
  Application of Semiconductor Detectors.

\end{thebibliography}
\bibliographystyle{spiebib} 

\end{document}